\documentclass{aa}

\usepackage{graphics}

\begin{document}

\thesaurus{04              
 (11.09.1 NGC 4945;  
  11.09.4;           
  11.19.3;           
  11.19.1)}          

\title{Mid-infrared ISO spectroscopy of NGC~4945
       \thanks{Based on observations with ISO, an ESA project with 
       instruments funded by ESA Member States (especially the PI 
       countries: France, Germany, the Netherlands and the United 
       Kingdom) and with the participation of ISAS and NASA}}

\author{H.W.W. Spoon\inst{1} \and J. Koornneef\inst{2} 
   \and A.F.M. Moorwood\inst{1} \and D. Lutz\inst{3}
   \and A.G.G.M. Tielens\inst{2}}

\offprints{H.W.W. Spoon}

\institute{European Southern Observatory,
           Karl-Schwarzschild Strasse 2, D-85748 Garching, Germany\\
           email: hspoon@eso.org, amoor@eso.org
       \and
           Kapteyn Astronomical Institute, Postbus 800, 
           NL-9700 AV Groningen, The Netherlands\\
           email: j.koornneef@astro.rug.nl, tielens@astro.rug.nl
       \and
           Max-Planck-Institut f\"ur extraterrestrische Physik (MPE),
           Postfach 1603, D-85740 Garching, Germany\\
           email: lutz@mpe.mpg.de}

\date{}

\maketitle

\begin{abstract}
We have observed the central region of the nearby starburst 
galaxy NGC~4945 with
the mid-infrared spectrometers SWS and PHT-S aboard ISO. We do not
find any evidence for the existence of the powerful AGN, inferred
from hard X-ray observations. The upper limits on our AGN tracers
$[\ion{Ne}{v}] $14.32$\mu$m\&24.3$\mu$m and $[\ion{Ne}{vi}]$ 7.65$\mu$m
imply an A$_V>$160 towards the NLR, assuming the NLR to be of equal
strength as in the Circinus galaxy. Other possibilities are discussed.
The starburst excitation indicators 
$[\ion{Ne}{iii}]$15.56$\mu$m/$[\ion{Ne}{ii}]$12.81$\mu$m and 
L$_{\rm bol}$/L$_{\rm lyc}$ suggest that the starburst in the central 
region is at least
5$\times$10$^6$ yrs old, and that it accounts for at least half of the 
nuclear bolometric luminosity. The starburst might well power the entire 
bolometric luminosity, but the available constraints are also consistent 
with an up to 50\% contribution of the embedded AGN. With PHT-S, 
at a resolution of $\approx$90, we detect strong absorption features 
of water ice, and, for the first time in an external galaxy, of CO$_2$
and CO. The same PHT-S spectrum
also reveals strong emission from the family of PAH features. Finally,
we have observed and detected several pure rotational and ro-vibrational
H$_2$ lines, two of which, the (0-0) S(0) \& S(1) lines, allow us to 
determine the excitation temperature (160K) and warm H$_2$ mass
(2.4$\times$10$^7$M$_{\odot}$). The low excitation temperature shows
Orion-like shocks not to be representative for the entire emission
of the central region of the galaxy and fairly normal PDRs to 
be perhaps more typical.

\end{abstract}

\section{Introduction}

NGC~4945 is a nearby, large (20$\arcmin\times$4$\arcmin$) spiral
galaxy seen nearly edge on (i $\sim$ 78$\degr$; Ott \cite{Ott}). 
At a recession velocity of 560km/s it is at the mean radial velocity
of the Centaurus group (Hesser et al. \cite{Hesser}), of which it is 
believed to be a member. Distance estimates vary between 3.5 and
4.0Mpc (see Bergman et al. \cite{Bergman} and Mauersberger et al. 
\cite{Mauersberger} for discussions). In this paper we will adopt a 
distance of 3.9 Mpc (Bergman et al. \cite{Bergman}), which implies 
that 1$\arcsec$ is equivalent to 18pc.

NGC~4945 is one of the brightest infrared galaxies in the sky:
S[12]=24Jy, S[25]=43Jy, S[60]=588Jy, S[100]=1416Jy (Rice et
al. \cite{Rice}). The total infrared luminosity amounts to 
L(8-1000$\mu$m)=2.95$\times$10$^{10}$ L$_{\sun}$, $\sim$75\% 
of which originates in the central 12$\arcsec\times$9$\arcsec$ 
(Brock et al. \cite{Brock}). 

Near infrared observations reveal the nuclear region to be the site
of a powerful, yet visually obscured, starburst. Br$\gamma$ (Moorwood 
et al. \cite{Moorwood96a}) and Pa$\alpha$ (Marconi et al. \cite{Marconi}) 
recombination line maps show the starburst to be concentrated in a 
circumnuclear disk or ring $\sim$200pc across (11$\arcsec$).
Further evidence for (a period of) strong star 
formation comes from the discovery of a conical structure, roughly 
perpendicular to the galaxy major axis. It is believed to be a cavity,
vacated by a starburst-driven superwind (Heckman et al. 
\cite{Heckman}; Moorwood et al. \cite{Moorwood96a}). The
non-detection of $[$\ion{O}{iii}$]$ within the cone and the absence 
of coronal lines excludes an AGN as the driver of the outflow.

Clear evidence for the presence of an AGN comes from hard X-ray
observations (Iwasawa et al. \cite{Iwasawa}; Guainazzi et al. 
\cite{Guainazzi}). The AGN X-ray emission is however heavily absorbed 
by a column density of 10$^{24.7}$cm$^{-2}$, which obscures the AGN at 
all optical and infrared wavelengths. Previous authors have attributed
most of the IR luminosity to the starburst (e.g. Moorwood \& Oliva 
\cite{Moorwood94}; Koornneef \& Israel \cite{Koornneef96}). Hard X-ray
observations with {\it BeppoSAX} indicate that the bolometric luminosity 
may as well be accounted for by the AGN alone (Guainazzi et al. 
\cite {Guainazzi}). 

3cm\&6cm ATCA radio maps of the central region of NGC~4945 
(Forbes \& Norris \cite{Forbes}) are dominated by strong nuclear 
emission, and emission extended along the disc of the galaxy. 
There is also evidence for some filamentry structure associated with 
the cavity cleared by the starburst superwind. VLBI observations by
Sadler et al. (\cite{Sadler}) reveal the existence of a compact radio
core. This, as well as the presence of H$_2$O megamasers in a Keplerian
disc about a $\sim$10$^6$ M$_{\odot}$ black hole (Greenhill et al. 
\cite{Greenhill}), are taken as further evidence for the presence 
of an AGN.

Near infrared observations of molecular hydrogen emission in
NGC~4945 have been reported by several authors over the last
15 years (e.g. Moorwood \& Glass \cite{Moorwood84}; Moorwood \& Oliva
\cite{Moorwood88}; Koornneef \cite{Koornneef93}; Moorwood \& Oliva
\cite{Moorwood94}; Koornneef \& Israel \cite{Koornneef96};
Moorwood et al. \cite{Moorwood96a}; Quillen et al. \cite{Quillen}; 
Marconi et al. \cite{Marconi}).
While fluxes are known for eight ro-vibrational transitions
accesible from the ground (Koornneef \& Israel \cite{Koornneef96}), 
spatial information is available only for the (1-0) S(1) 2.1218$\mu$m 
line. These observations show the H$_2$ emission to be associated with
the hollow cone, {\it not} with the starburst traced in hydrogen 
recombination emission. The absence of a correlation argues against 
photons as the source of excitation. Instead, the emission 
is attributed to shock heating of the molecular material at the
face of the cavity (Moorwood et al. \cite{Moorwood96a}; Marconi et
al. \cite{Marconi}). 

Mid-infrared spectroscopy is much less affected by intervening 
extinction than the UV and optical equivalents, with 
A($\lambda$)/A$_V$ less than 0.1. Observations of the central region 
of the galaxy, using the mid-infrared spectrometer 
SWS (De Graauw et al. \cite{deGraauw}) and the spectrophotometer 
PHT-S (Lemke et al. \cite{Lemke}), both aboard ISO (Kessler et al. 
\cite{Kessler}), are therefore very useful to study the nuclear 
components otherwise hidden by heavy extinction. In Sect.~3.1 
we present the results of the search for high excitation emission 
from the AGN. In Sect.~3.2 we study the properties of the nuclear
starburst. In Sect.~3.3 we discuss the dominant nuclear power 
source. Sect.~3.4 discusses the broad emission and absorption 
features, tracing the properties of the interstellar medium in and 
in front of the nucleus. Finally, in Sect.~3.5 we discuss the 
physical conditions and excitation of the warm molecular hydrogen.

\section{Observations}

As part of the Central Program ``MPEXGAL'', we have
observed the central region of NGC~4945 with the Short Wavelength 
Spectrometer (SWS) and the spectrophotometer PHT-S on board ISO.

\subsection{SWS spectroscopy}

SWS grating line profile scans (SWS02 mode) were obtained on 1996
February 6 for 28 spectral lines in the range of 2.42 to 
40.34$\mu$m. The spectral resolution in this range varies between 
R=900 and 2000, corresponding to a velocity resolution of 330--150 km/s.
Aperture sizes used range between 14$\arcsec\times$20$\arcsec$ and 
20$\arcsec\times$33$\arcsec$. SWS was centered on the 1.4GHz continuum
peak from Ables et al. (\cite{Ables}), which coincides with the the 
position of the L-band peak of Moorwood et al. (\cite{Moorwood96a}). 
At the time of observation the position angle of the major axis of the
SWS apertures was 35.4$\degr$, 10$\degr$ off from that of the galaxy
major axis (45$\degr$).

\begin{figure*}
\resizebox{\hsize}{!}{\includegraphics{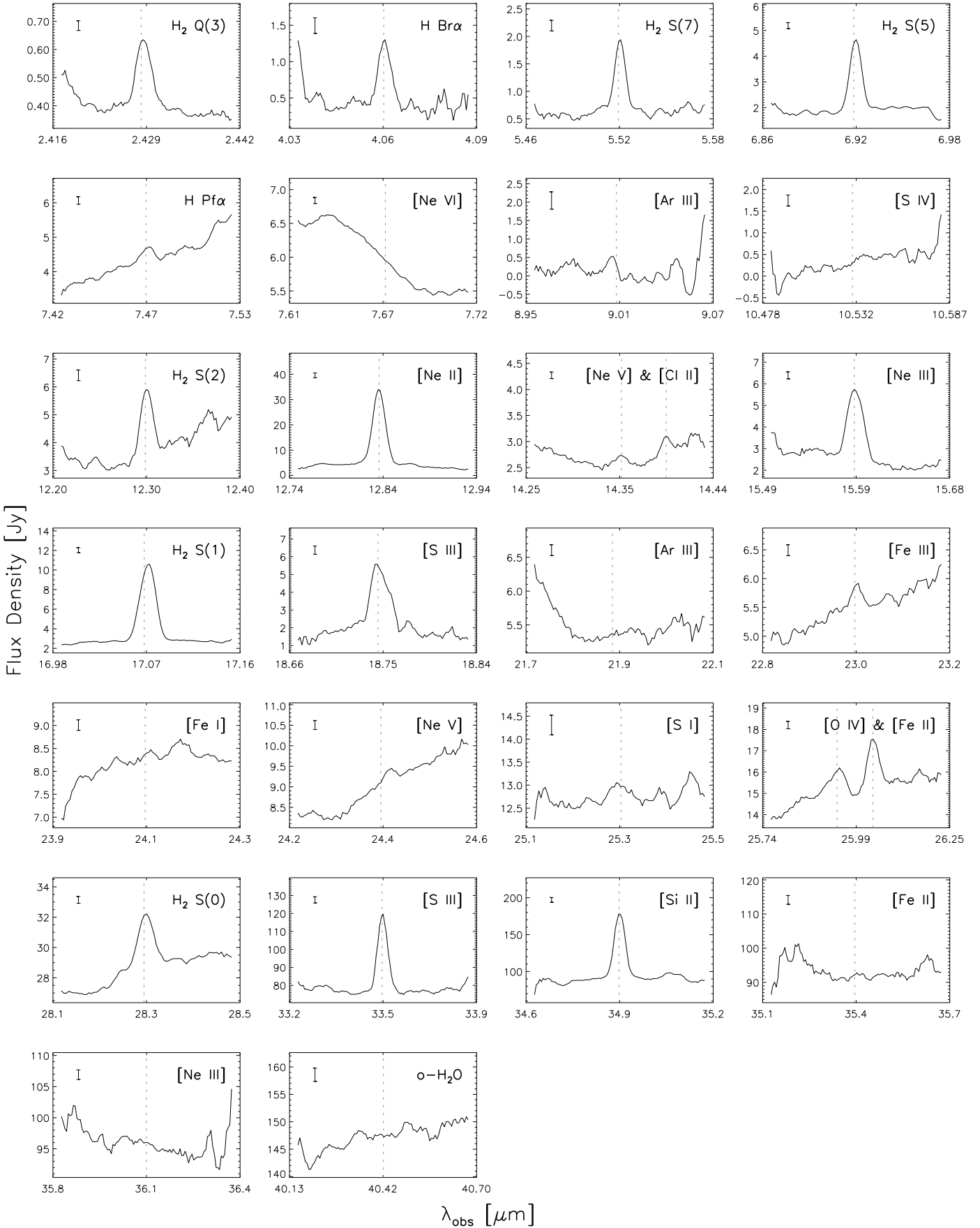}}
\caption{SWS line spectra of NGC~4945. Typical $\pm$1$\sigma$ error bars 
are marked. Note that noise increases towards the edges of scans}
\label{swsspec}
\end{figure*}

The data reduction was performed using the SWS Interactive Analysis 
system (SIA; Lahuis et al. \cite{Lahuis}; Wieprecht et
al. \cite{Wieprecht}), putting special emphasis on tools to improve
cosmic ray spike removal, dark current subtraction and flat fielding. 
The wavelength calibration of SWS is discussed by Valentijn et al. 
(\cite{Valentijn96a}). We used calibration files as of March 1999.
The accuracy of the flux calibration is estimated to be 30\% (Schaeidt
et al. \cite{Schaeidt}). The resulting spectra are shown in 
Fig~\ref{swsspec}.

In total 17 spectral lines were detected. For another 
11 lines we derived upper limits, using gaussian profiles 
of width equal to other lines of the same (or comparable) species, 
scaled to a peak height corresponding to approximately 3$\sigma$ of 
the noise. Both detections and upper limits are presented in 
Table~\ref{swsfluxes}.

\begin{table}
\caption[]{NGC~4945 results from SWS observations}
\begin{tabular}{llll} \\ \hline
Identification & $\lambda_{\rm rest}$ & F$_{\rm obs}$ & Aperture \\ 
               & $[\mu$m$]$ & $[$W/cm$^{2}]$  & $[\arcsec\times\arcsec]$ \\ \hline
H$_2$ (1-0) Q(3)        &  2.424 & 3.20 10$^{-20}$   & 14$\times$20 \\
$\ion{H}{i}$ Br$\alpha$ &  4.052 & 7.79 10$^{-20}$   & 14$\times$20 \\
H$_2$ (0-0) S(7)        &  5.510 & 1.11 10$^{-19}$   & 14$\times$20 \\
H$_2$ (0-0) S(5)        &  6.909 & 1.54 10$^{-19}$   & 14$\times$20 \\
$\ion{H}{i}$ Pf$\alpha$ &  7.460 & 1.95 10$^{-20}$   & 14$\times$20 \\
$[\ion{Ne}{vi}]$        &  7.652 & $<$7.3 10$^{-21}$ & 14$\times$20 \\
$[\ion{Ar}{iii}]$       &  8.991 & $<$2.3 10$^{-20}$ & 14$\times$20 \\
$[\ion{S}{iv}]$         & 10.511 & $<$8.8 10$^{-21}$ & 14$\times$20 \\
H$_2$ (0-0) S(2)        & 12.279 & 7.45 10$^{-20}$   & 14$\times$27 \\
$[\ion{Ne}{ii}]$        & 12.814 & 8.68 10$^{-19}$   & 14$\times$27 \\
$[\ion{Ne}{v}]$         & 14.322 & $<$7.0 10$^{-21}$ & 14$\times$27 \\
$[\ion{Cl}{ii}]$        & 14.368 & 5.19 10$^{-21}$   & 14$\times$27 \\
$[\ion{Ne}{iii}]$       & 15.555 & 7.51 10$^{-20}$   & 14$\times$27 \\
H$_2$ (0-0) S(1)        & 17.035 & 1.51 10$^{-19}$   & 14$\times$27 \\
$[\ion{S}{iii}]$        & 18.713 & 6.88 10$^{-20}$   & 14$\times$27 \\
$[\ion{Ar}{iii}]$       & 21.829 & $<$5.1 10$^{-21}$ & 14$\times$27 \\
$[\ion{Fe}{iii}]$       & 22.925 & 6.46 10$^{-21}$   & 14$\times$27 \\
$[\ion{Fe}{i}]$         & 24.042 & $<$5.4 10$^{-21}$ & 14$\times$27 \\
$[\ion{Ne}{v}]$         & 24.318 & $<$5.5 10$^{-21}$ & 14$\times$27 \\
$[\ion{S}{i}]$          & 25.257 & $<$1.4 10$^{-20}$ & 14$\times$27 \\
$[\ion{O}{iv}]$         & 25.890 & 3.00 10$^{-20}$   & 14$\times$27 \\
$[\ion{Fe}{ii}]$        & 25.988 & 4.11 10$^{-20}$   & 14$\times$27 \\
H$_2$ (0-0) S(0)        & 28.221 & 4.82 10$^{-20}$   & 20$\times$27 \\
$[\ion{S}{iii}]$        & 33.481 & 4.87 10$^{-19}$   & 20$\times$33 \\
$[\ion{Si}{ii}]$        & 34.815 & 9.66 10$^{-19}$   & 20$\times$33 \\
$[\ion{Fe}{ii}]$        & 35.349 & $<$2.8 10$^{-20}$ & 20$\times$33 \\
$[\ion{Ne}{iii}]$       & 36.014 & $<$1.6 10$^{-20}$ & 20$\times$33 \\
o-H$_2$O                & 40.341 & $<$2.8 10$^{-20}$ & 20$\times$33 \\ \hline
\end{tabular}
\label{swsfluxes}
\end{table}

\subsection{PHT-S spectrophotometry}

We have obtained two low resolution ($\lambda$/$\Delta\lambda\sim$90) 
ISO-PHT-S spectra, on 1996 October 12 and 1997 August 3, respectively.
ISO-PHT-S comprises two low-resolution grating spectrometers covering
simultaneously the wavelength range 2.47 to 4.87$\mu$m and 5.84 to
11.62$\mu$m. The spectrum is registered by two linear arrays of 64
Si:Ga detectors with a common entrance aperture of 24$\arcsec$ $\times$
24$\arcsec$.
The measurements were carried out in rectangular chopped mode, using
a chopper throw of 180$\arcsec$. The resulting spectra thus are free of
contributions from zodiacal light, that would otherwise affect the 
spectrum. The pure on-source integration times were 512 and 1024 s.

The ISO-PHT-S data were reduced using PIA\footnote{PIA is a joint 
development by the ESA Astrophysics Division and the ISO-PHT Consortium} 
(Gabriel et al. \cite{Gabriel}) version 8.1. Steps in the data 
reduction included: 1) 
deglitching on ramp level. 2) subdivision of ramps in two sections 
of 32 non destructive read-outs. 3) ramp fitting to derive 
signals. 4) masking of bad signals by eye-inspection. 5) kappa sigma 
and min/max clipping on remaining signal distribution. 6)
determination of average signal per chopper plateau. 7) masking or 
correction of bad plateaux by eye-inspection. 8) background
subtraction using all but the first four plateaux. 9) finally, flux
calibration, using the signal dependent spectral response function of 
Acosta-Pulido (\cite{Acosta99}). This spectral response function
avoids some deficiencies of the previous PIA response function, in 
particular in the 3$\mu$m region near the ``Ice'' feature. The 
absolute calibration is accurate to within 20\%.

\begin{figure*}
\resizebox{\hsize}{10cm}{\includegraphics{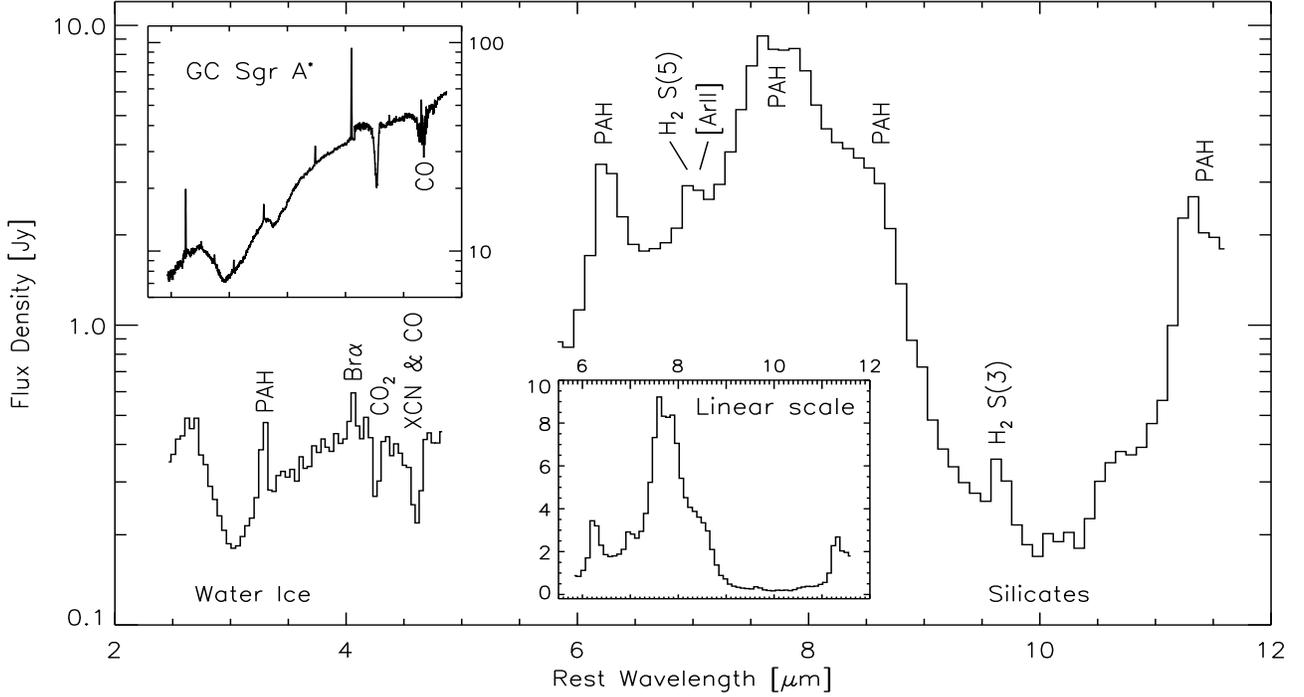}}
\caption{The average ISO-PHT-S spectrum of NGC~4945. Upper inset: 
comparison with the ISO-SWS spectrum of the line of sight towards 
the Galactic center. Lower inset: the ISO-PHT-SL spectrum in linear 
flux scale}
\label{phtspec}
\end{figure*}

The two resulting spectra were obtained at slightly different position
angles about the nucleus. For the first, the square aperture was
aligned with the galaxy major axis (45$\degr$). For the second, the 
position angle was 31.1$\degr$. 
Fig.~\ref{phtspec} shows the averaged ISO-PHT-S spectrum. The on-source 
integration times were used as weight factors in the computation of
the average spectrum.

A number of emission lines can be identified in the ISO-PHT-S
spectrum. These include 9.66$\mu$m H$_2$ (0-0) S(3), the
unresolved blend of 6.99$\mu$m $[$\ion{Ar}{ii}$]$ and 6.91$\mu$m
H$_2$ (0-0) S(5), and 4.05$\mu$m H Br$\alpha$. For the 9.66$\mu$m
H$_2$ (0-0) S(3) line, not included in the SWS02 line scans, we
measure a flux of 5.4$\times$10$^{-20}$ W/cm$^{2}$, with an
uncertainty of 30\%.

\section{Results}

\subsection{AGN not seen at mid-infrared wavelengths}

NGC~4945 is a peculiar and interesting target for studying the
relation of AGN and star formation in galaxies. Clear evidence 
for hidden AGN activity comes from hard X-ray observations.
NGC~4945 is amongst the brighest hard X-ray emitting galaxies
and exhibits variability of its 13--200keV flux on timescales
of $\sim$10hrs, which clearly establishes its AGN origin 
(Iwasawa et al. \cite{Iwasawa}; Guainazzi et al. 
\cite{Guainazzi}; Marconi et al. \cite{Marconi}). The AGN X-ray 
emission is heavily absorbed by a column density of 10$^{24.7}$ 
cm$^{-2}$ (corresponding to A$_V$$\sim$2600), a high 
value, but within the range observed for Seyfert 2 galaxies 
(e.g. Risaliti et al. \cite{Risaliti}). In unified
schemes, the X-ray obscuration measures a line of sight towards
the very center. Obscuration towards the NLR probes a different
line of sight and is usually significantly lower, making the NLR
visible in Seyfert 2 galaxies.

Mid-infrared high excitation lines are able to penetrate a far 
larger dust obscuration than their optical and UV counterparts.
They are therefore ideally suited as tracers of embedded AGN 
activity. Mid-infrared emission lines like $[\ion{Ne}{v}]$ 
14.32$\mu$m \& 24.32$\mu$m, $[\ion{Ne}{vi}]$ 7.65$\mu$m and 
$[\ion{O}{iv}]$ 25.9$\mu$m are prominently present in the spectrum 
of all Seyferts observed with ISO (Moorwood et al. \cite{Moorwood96b};
Genzel et al. \cite{Genzel}). On the other hand, the same emission lines 
are also weakly visible towards, for instance, supernova remnant 
RCW~103 (Oliva et al. \cite{Oliva}). $[\ion{O}{iv}]$ emission 
(at the few percent level compared to $[\ion{Ne}{ii}]$) has also 
been detected in a sample of starburst galaxies (Lutz et al. 
\cite{Lutz98}), again at a level much weaker than seen in typical 
AGNs. The origin of the weak level emission in these sources is 
believed to be shocks. A detection of any of the high excitation 
lines discussed above does therefore not automatically imply the 
detection of an AGN in NGC~4945.

We do not detect the lines of $[\ion{Ne}{v}]$ and $[\ion{Ne}{vi}]$. 
No trace of $[\ion{Ne}{vi}]$ is seen in the wing of the nearby PAH 
emission feature (Fig.~\ref{swsspec}). From Fig.~\ref{swsspec} it 
might appear that the two $[\ion{Ne}{v}]$ lines were indeed detected. 
However, at the level at which the features appear, instrumental 
effects play a significant role. In the 14.35$\mu$m line scan a strong
fringe in the relative spectral response function coincides 
{\it exactly} with the expected position for the $[\ion{Ne}{v}]$ line. 
Depending on the size of the emitting area, the feature may be 
entirely attributed to this instrumental effect. We therefore chose 
to state an upper limit for the 14.32$\mu$m $[\ion{Ne}{v}]$ line. 
The feature seen in the other $[\ion{Ne}{v}]$ scan, centered at 
24.4$\mu$m, was registered by only two detectors, although 12 detectors
scanned over the central wavelength. As visible in Fig.~\ref{swsspec},
the feature is redshifted with respect to the NGC~4945 systemic velocity.
This shift is not observed for any other line we observed. We
therefore derive an upper limit for this $[\ion{Ne}{v}]$ line too.

The only detected high ionization line in NGC~4945 is the 25.9$\mu$m 
$[\ion{O}{iv}]$ line. An AGN contribution to this line is possible 
--- to match the limits on higher excitation lines, only part of the 
$[\ion{O}{iv}]$ emission would be related to an AGN. The detection of 
possibly shock-related $[\ion{O}{iv}]$ in many starbursts (Lutz et al. 
\cite{Lutz98}) cautions, however, that this may be a more plausible 
origin of $[\ion{O}{iv}]$ in NGC~4945. The ratio of 0.033 with respect
to $[\ion{Ne}{ii}]$+0.44$\times[\ion{Ne}{iii}]$ is above average, but 
well within the range observed for the Lutz et al. (\cite{Lutz98}) 
starbursts, also considering that the high extinction (Sect.~3.2) 
will increase the observed ratio relative to the intrinsic one. 
A population of Wolf-Rayet stars as origin of the $[\ion{O}{iv}]$
emission seems unlikely. Lutz et al. (\cite{Lutz98}) have shown that
the $[\ion{O}{iv}]$ emission would have to originate in widely
dispersed small \ion{H}{ii} regions and would have to be relatively 
strong. $[\ion{O}{iv}]$ emission at this level has not been observed
in local star forming regions. A conservative analysis will hence not
attribute the $[\ion{O}{iv}]$ emission in NGC~4945 to the AGN nor
to a population of Wolf-Rayet stars.

\begin{table}
\caption[]{In X-ray properties NGC~4945 is very similar to Circinus. 
However, IR indicators for a narrow line region are missing in NGC~4945}
\begin{tabular}{lll} \\ \hline
                                          & NGC~4945 & Circinus \\ \hline
Distance                $[$Mpc$]$         & 3.9 & 4$^{\mathrm{a}}$ \\
H col. dens. to AGN       $[$cm$^{-2}$$]$   & 5$\times$10$^{24}$ & 4$\times$10$^{24}$$^{\mathrm{b}}$ \\
A$_V$ to AGN                              & 2600 & 2100 \\
L(2--10keV)             $[$L$_{\odot}$$]$ & 78$\times$10$^7$$^{\mathrm{c}}$ & 8.8--44$\times$10$^7$$^{\mathrm{b}}$ \\ 
$\nu$L$_{\nu}$(100keV)/L$_{\rm FIR}$      & 0.003$^{\mathrm{d}}$ & 0.002$^{\mathrm{d}}$ \\
L(8--1000$\mu$m)        $[$L$_{\odot}$$]$ & 2.2$\times$10$^{10}$ & 1.2$\times$10$^{10}$$^{\mathrm{a}}$ \\ 
$[\ion{Ne}{v}]$14.3$\mu$m $[$W/cm$^2$$]$  & $<$7.0$\times$10$^{-21}$ &440$\times$10$^{-21}$$^{\mathrm{e}}$\\
$[\ion{Ne}{v}]$24.3$\mu$m $[$W/cm$^2$$]$  & $<$5.5$\times$10$^{-21}$ &244$\times$10$^{-21}$$^{\mathrm{e}}$\\
$[\ion{Ne}{vi}]$7.65$\mu$m $[$W/cm$^2$$]$ & $<$7.3$\times$10$^{-21}$ &413$\times$10$^{-21}$$^{\mathrm{e}}$\\ \hline
\end{tabular}
\label{N4945CircinusComparison}
\begin{list}{}{}
\item[$^{\mathrm{a}}$] Siebenmorgen et al. (\cite{Siebenmorgen})
\item[$^{\mathrm{b}}$] Matt et al. (\cite{Matt})
\item[$^{\mathrm{c}}$] Guainazzi et al. (\cite{Guainazzi})
\item[$^{\mathrm{d}}$] Marconi et al. (\cite{Marconi})
\item[$^{\mathrm{e}}$] Moorwood et al. (\cite{Moorwood96b})
\end{list}
\end{table}

The limits on high excitation AGN tracers are consistent with several
scenarios, or perhaps more likely a combination of them:
\begin{itemize}
\item {\bf The Narrow Line Region is extremely obscured even in the mid-IR.}
      We derive an A$_{V}\geq$160 
      (A(7.65$\mu$m)=A(14.3$\mu$m)=A(24.3$\mu$m)$\geq$4.3) to the NLR
      from a comparison of Circinus and NGC~4945 $[\ion{Ne}{v}]$ and 
      $[\ion{Ne}{vi}]$ line strengths, under the assumption that the galaxies'
      NLRs are similar. The choice for Circinus is motivated in 
      Table~\ref{N4945CircinusComparison}.
\item {\bf UV photons from the AGN are absorbed close to the nucleus along
      all lines of sight}
\item {\bf The extreme ultraviolet luminosity of the AGN is lower than in 
      Circinus.} In comparison to the Circinus SED, this would imply a large
      deficiency in UV relative to X-ray flux
      (Table~\ref{N4945CircinusComparison}).
\end{itemize}

\subsection{Starburst properties}

Near-infrared broad-band and emission-line imaging has revealed 
the nucleus of NGC~4945 to be the site of a sizeable starburst,
the presence of which is illustrated by the conically 
shaped starburst superwind-blown cavity traced at many near-infrared 
wavelengths (Moorwood et al. \cite{Moorwood96a}; Marconi et al. 
\cite{Marconi}). Hampered by the large extinction even in the
near-infrared, age estimates for the nuclear starburst are sparse
and intrinsically uncertain. ISO-SWS offers the possibility for the 
first time to observe the mid-infrared line ratio $[\ion{Ne}{iii}]$
15.56$\mu$m/$[\ion{Ne}{ii}]$ 12.81$\mu$m. This ratio, which
is much less affected by extinction than visible and UV lines, 
is sensitive to the hardness of the stellar radiation field
and hence is a good indicator for the age of the nuclear starburst.
We observed the two lines in the same ISO-SWS aperture, which was
centered on the nucleus (see Table~\ref{swsfluxes}).

To estimate the extinction to the NGC~4945 nuclear starburst we use the 
ratio of the 18.71$\mu$m and 33.48$\mu$m $[$\ion{S}{iii}$]$ lines. This 
ratio is commonly used as a density diagnostic for the density range
10$^{2.5}$--10$^{4.5}$ cm$^{-3}$ and is only mildly dependent on the
temperature of the emitting gas. Assuming a typical starburst gas
density of 300 cm$^{-3}$ (Kunze et al. \cite{Kunze}; Rigopoulou et
al. \cite{Rigopoulou96}), the intrinsic ratio should be $\sim$0.43 
(i.e. the value in the low density limit, computed using the collision
strengths of Tayal \& Gupta \cite{Tayal}). The observed ratio is far 
lower: 0.14$\pm$0.06. We hence deduce a screen extinction of 
A(18.7$\mu$m)=1.7$^{+0.8}_{-0.5}$, which, using the Galactic center 
extinction law of Draine (\cite{Draine}; with A(9.7$\mu$m)/E(J--K)=0.71), 
amounts to A$_V$=36$^{+18}_{-11}$. This value is to be considered an
upper limit in case the $[$\ion{S}{iii}$]$ emitting area is larger than 
14$\arcsec\times$27$\arcsec$, in which case an aperture effect causes
the $[$\ion{S}{iii}$]$ ratio to be a lower limit.

Another independent estimate of the extinction is usually obtained
from hydrogen recombination line strengths, assuming `case-B'
conditions. For NGC~4945 we therefore observed the 4.05$\mu$m Br$\alpha$
and the 7.46$\mu$m Pf$\alpha$ line. Both were measured in the same
aperture. The ratio Pf$\alpha$/Br$\alpha$ is 0.25$\pm$0.10, whereas
`case-B' recombination theory predicts a ratio of 0.32. The extinction 
at 7.46$\mu$m must therefore be similar or slightly {\it larger} than at 
4.05$\mu$m. This indicates that the grain composition is unusual and 
probably more similar to the composition found in the line of sight 
towards the Galactic center (Lutz et al. \cite{Lutz96}; Lutz \cite{Lutz99})
than found towards other parts of our Galaxy. An extinction towards
the NGC~4945 nuclear starburst can therefore at present not be derived
from lines in the 4--7$\mu$m range.

The extinction derived for the nuclear starburst is somewhat larger than
the value we derive for the warm molecular hydrogen (see Sect.~3.5). 
This indicates that the warm molecular hydrogen and nuclear starburst 
emission are coming from different nuclear components, the latter 
possibly more enshrouded than the former. With the unusual grain 
composition in mind, it is striking how well the Galactic Center
extinction law fits our molecular hydrogen data, resulting in a smooth
excitation diagram, even for the H$_2$ (0-0) S(3) line in the center
of the 9.7$\mu$m silicate feature (see Sect.~3.5). We are therefore 
confident that the extinction correction for the starburst, derived 
using the $[$\ion{S}{iii}$]$ ratio, is also useful.

In order to determine the excitation of the nuclear starburst we apply
the extinction correction derived from the $[\ion{S}{iii}]$ ratio to the 
observed $[\ion{Ne}{iii}]$/$[\ion{Ne}{ii}]$ ratio. The extinction
corrections amount to A(12.8$\mu$m)=1.51 and A(15.6$\mu$m)=1.19. 
The extinction corrected $[\ion{Ne}{iii}]$/$[\ion{Ne}{ii}]$ ratio is 
0.064$^{+0.037}_{-0.032}$. Thornley et al. (\cite{Thornley}) list 
observed $[\ion{Ne}{iii}]$/$[\ion{Ne}{ii}]$ ratios for 26 starburst 
galaxies, all measured in the same ISO-SWS configuration. Clearly, 
NGC~4945 is among the lowest excitation targets in their sample.
Note that the ISO-SWS aperture used is large compared to the 
typical size scales in starbursts. The ratios listed by Thornley et
al. (\cite{Thornley}) should therefore be regarded as aperture averaged.

For starburst galaxies L$_{\rm bol}$/L$_{\rm lyc}$ is another measure of the 
excitation of star clusters. Depending on the upper mass cut-off, the
star formation decay time scale and the age of the clusters, Thornley 
et al. (\cite{Thornley}) modeled L$_{\rm bol}$/L$_{\rm lyc}$ to lie between
3 and 200. The measured values for starburst galaxies range between 3 
and 50. Below we will determine L$_{\rm bol}$/L$_{\rm lyc}$ for the NGC~4945 
nuclear starburst. We assume L$_{\rm bol}$=L$_{\rm IR}$ (i.e. {\it no}
AGN contribution to L$_{\rm IR}$) and estimate L$_{\rm lyc}$ from
the dereddened 4.05$\mu$m Br$\alpha$ flux. For A(4.05$\mu$m)=1.2 
(applying the Galactic center law of Draine (\cite{Draine}) for 
A$_V$=36$^{+18}_{-11}$) and L$_{\rm lyc}$=670 L$_{\rm Br\alpha}$ we find 
L$_{\rm lyc}$=8$^{+9}_{-4}$$\times$10$^{8}$L$_{\odot}$ and 
L$_{\rm bol}$/L$_{\rm lyc}$=28$^{+26}_{-15}$. Using the 12.81$\mu$m
$[\ion{Ne}{ii}]$ line and the empirical scaling L$_{\rm lyc}$=64 
L$_{\ion{Ne}{ii}}$ (Genzel et al. \cite{Genzel}) a similar result is 
obtained. 

Given the variety of possible star forming histories, it is hard to 
constrain the age of the nuclear starburst (assuming {\it no} AGN 
contribution to L$_{\rm IR}$). However, both excitation diagnostics agree 
on a low excitation which suggests an evolved burst with an age in 
excess of 5$\times$10$^{6}$ years, but would also be consistent with 
a low IMF upper mass cut-off.

Marconi et al. (\cite{Marconi}) show that it is possible
to construct starburst models which are consistent with their 
near-infrared observations of NGC~4945, but differ by the total 
luminosity generated (their Fig.~4). An instantaneous burst would
not leave space in the energy budget for a sizable contribution 
from the hidden AGN, whereas a combination of instantaneous burst
and constant star formation would. We would like to point out here
that the latter model would be inconsistent with the low
$[\ion{Ne}{iii}]$/$[\ion{Ne}{ii}]$ ratio observed by us. Only
their instantaneous burst is in agreement with both the near-infrared
and mid-infrared observations.

\subsection{What powers the nucleus of NGC~4945?}

The large extinction towards the nuclear starburst and the AGN buried within,
makes it very difficult to assess the contributions of either component to the 
nuclear bolometric luminosity.\\
The optical, near-infrared, mid-infrared and far-infrared spectra of NGC~4945
are entirely consistent with a starburst-like nature: BLR or NLR 
high-excitation lines are absent; the starburst excitation indicator 
$[\ion{Ne}{iii}]$/$[\ion{Ne}{ii}]$ has a starburst-like value; the ratios 
of 6$\mu$m (ISO-PHT-S), S12 or S25 to S60 or S100 fluxes are all very low 
and consistent with emission from cold dust only. Furthermore, the ratio 
L$_{\rm bol}$/L$_{\rm lyc}$=28$^{+26}_{-15}$, is perfectly consistent with the
low excitation of the starburst as deduced from 
$[\ion{Ne}{iii}]$/$[\ion{Ne}{ii}]$. And last, NGC~4945 lies on the
radio-far-infrared correlation for starburst galaxies (Forbes \& Norris
\cite{Forbes}). Hence, the starburst might well account for the the
entire observed bolometric luminosity.

On the other hand, Guainazzi et al. (\cite{Guainazzi}), who have
observed the AGN in NGC~4945 in hard X-rays, compute the AGN to be able
to account for all the bolometric luminosity observed, {\it if} it has
a typical quasar L$_{\rm X}$/L$_{\rm bol}$ ratio. Since there is no 
such thing as a template AGN spectrum, the conversion factor applied, 
L$_{\rm 1-10keV}$/L$_{\rm bol}\sim$0.05 (Elvis et al. \cite{Elvis}), may have
an uncertainty which could easily allow for the NGC~4945 starburst to 
dominate the bolometric luminosity instead.\\
The same uncertanties surround the accretion rate of the 
$\sim$1.6$\times$10$^6$M$_{\odot}$ black hole inferred from H$_2$O
maser observations (Greenhill et al. \cite{Greenhill}). A high but not
implausible rate of 50\% of the Eddington rate
(L$_{\rm Edd}$$\sim$4.1$\times$10$^{10}$ L$_{\odot}$)
would suffice to power the observed bolometric luminosity. Given the 
wide range of efficiencies inferred for Seyferts, this information does 
not add anything to identify the dominant power source.

In this complex situation with two potentially dominant power sources,
the most significant constraint on their relative weight is the
total L$_{\rm bol}$/L$_{\rm lyc}$ ratio and its implications on the 
L$_{\rm bol}$/L$_{\rm lyc}$ of the {\em starburst} component. 
L$_{\rm lyc}^{\rm sb}$ is directly constrained by observations, but 
L$_{\rm bol}^{\rm sb}$ changes for different assumptions on the 
starburst and AGN contributions to the total bolometric luminosity. 
If there is a significant AGN contribution, 
(L$_{\rm bol}$/L$_{\rm lyc}$)$_{\rm sb}$ will be less than the global 
value of 28. Values as low as $\sim$3 which are possible for a zero
age massive star population with Salpeter IMF (e.g. Leitherer \& Heckman 
\cite{Leitherer}) are strongly inconsistent with the low excitation 
observed in NGC~4945. Thornley et al. (\cite{Thornley}) model 
$[$\ion{Ne}{iii}$]$/$[$\ion{Ne}{ii}$]$ and L$_{\rm bol}$/L$_{\rm lyc}$ ratios 
of starbursts, taking into account clusters of different ages
within the ISO-SWS aperture. An evolving starburst with
$[$\ion{Ne}{iii}$]$/$[$\ion{Ne}{ii}$]$=0.064 as in NGC~4945 must have 
a L$_{\rm bol}$/L$_{\rm lyc}$$\succeq$15 (their Fig.~8). This limit
simply reflects the higher L$_{\rm bol}$/L$_{\rm lyc}$ of later type O stars and
persists if the low excitation is due to an upper mass cut-off rather
than evolution. With a lower limit of $\sim$15 on 
(L$_{\rm bol}$/L$_{\rm lyc}$)$_{\rm sb}$, the starburst contribution to the 
bolometric luminosity must be at least $\sim$50\%.

We hence conclude that the AGN in NGC~4945 plays a secondary although most
likely not insignificant role in the energetics of this nearby galaxy.
Extremely small values for the AGN contribution to the bolometric
luminosity would imply an unrealistically high ratio of 
L$_{\rm X}$/L$_{\rm bol}$ for the AGN. The very low inferred black
hole mass, the very cold mid-infrared to far-infrared colors, and the 
absence of any clear line of sight towards the AGN, support our view that 
starburst activity dominates AGN activity in NGC~4945.

\subsection{Emission and absorption features}

The infrared spectrum of the central region of NGC~4945 obtained with
ISO-PHT-S (see Fig.~\ref{phtspec}) presents a new view of the ISM
in starburst galaxies. Even at the low spectral resolving power of 
R$\approx$90 the spectrum is dominated by a wealth of emission and 
absorption features. 

Especially prominent is the family of PAH emission features at 3.3, 
6.2, 7.7, 8.6 and 11.3$\mu$m, which ISO confirmed to be common-place 
in most galactic and extragalactic ISM spectra (e.g. Acosta-Pulido et al. 
\cite{Acosta96}; Rigopoulou et al. \cite{Rigopoulou99}; Mattila et al.
\cite{Mattila}; Clavel et al. \cite{Clavel}). 
Nevertheless, the weakness of the 8.6 and 11.3$\mu$m PAH bands 
in NGC~4945 is unusual. Consistent with A$_V$$\sim$36 and with 
the strength of the absorption features discussed below, we 
explain this weakness by heavy extinction, which will 
suppress these two features placed in the wings of the silicate 
absorption feature.

Perhaps the most important result, however, is the rich absorption 
spectrum, indicating that we are observing the infrared sources in 
the central region of NGC~4945 through a medium containing molecular 
ices. Interstellar absorptions of 4.27$\mu$m (2343cm$^{-1}$) solid 
CO$_2$ and 4.68--4.67$\mu$m `XCN'+CO are detected, the first 
time in an extragalactic source to our knowledge. 
At our resolving power and signal-to-noise we cannot determine
the contribution of 4.62$\mu$m (2165cm$^{-1}$) XCN, 4.67$\mu$m 
(2140cm$^{-1}$) CO ice and of gaseous CO absorptions to the 
4.58--4.67$\mu$m absorption complex. The strength of the XCN 
absorption appears to be remarkable, suggestive of ice grain processing 
in an energetic environment (Lacy et al. \cite{Lacy}). We defer 
a more detailed analysis of the XCN/CO feature to a future paper, 
which will also include the results of follow-up observations 
with ISAAC at the VLT. A strong silicate feature is observed around 
9.7$\mu$m (see also Moorwood \& Glass \cite{Moorwood84}). A deep 
minimum is also detected around 3.0$\mu$m, which is suggestive of 
water ice (or more precise, the O-H stretch) absorption. 
Table~\ref{coldens} gives column densities for some of the 
absorption features discussed above. The presence and strength of
these absorption features is consistent with the high starburst 
obscuration derived from the emission lines (but see also Chiar et al.
(\cite{Chiar}) for variations in the strength of features along lines
of sight of similar A$_V$).

\begin{table*}
\caption[]{Column densities of solid state features towards the
nucleus of NGC~4945 and towards the Galactic center (SgrA$^*$).
In order to derive the column densities we integrated 
$\int\tau_{\nu}$d$\nu$ over the width of the band and divided the
result by the bandstrength A. N$_{\rm H}$ was determined from 
N$_{\rm H}$=1.9$\times$10$^{21}$ A$_V$, where A$_V$=30 for SgrA$^*$ and
A$_V$=36 for NGC~4945}
\begin{tabular}{llccccccc} \\ \hline
Species & $\lambda_{\rm rest}$ & A & \multicolumn{2}{c}{$\tau$$_{\rm center}$} & \multicolumn{2}{c}{N} & \multicolumn{2}{c}{N/N$_{\rm H}$}\\ 
  & $[\mu$m$]$ & $[$cm/mol.$]$ &     &     & \multicolumn{2}{c}{[mol./cm$^{2}]$} & \multicolumn{2}{c}{}\\
 & & & NGC~4945 & SgrA$^*$ & NGC~4945 & SgrA$^*$ & NGC~4945 & SgrA$^*$\\ \hline
H$_2$O&3.09&  2$\times$10$^{-16}$$^{\mathrm{a}}$&0.9&0.50$^{\mathrm{b}}$&2.4$\times$10$^{18}$&1.3$\times$10$^{18}$$^{\mathrm{b}}$&3.5$\times$10$^{-5}$&2.3$\times$10$^{-5}$\\
CO$_2$&4.27&7.4$\times$10$^{-17}$$^{\mathrm{c}}$&0.8$^{\mathrm{d}}$&0.72$^{\mathrm{e}}$&2.0$\times$10$^{17}$&1.7$\times$10$^{17}$$^{\mathrm{e}}$&2.9$\times$10$^{-6}$&3.0$\times$10$^{-6}$\\
XCN   &4.60&---&---&---&---&---&---&---\\ \hline
\end{tabular}
\label{coldens}
\begin{list}{}{}
\item[$^{\mathrm{a}}$] Hagen \& Tielens (\cite{Hagen})
\item[$^{\mathrm{b}}$] Chiar et al. (\cite{Chiar})
\item[$^{\mathrm{c}}$] Gerakines et al. (\cite{Gerakines95})
\item[$^{\mathrm{d}}$] Determined by fitting a Gaussian profile
                       followed by rebinning to the ISO-PHT-S instrument resolution
\item[$^{\mathrm{e}}$] Gerakines et al. (\cite{Gerakines99})
\end{list}
\end{table*}

At the resolution of ISO-PHT-S the molecular absorption features 
in NGC~4945 show striking similarities with the features seen in the
ISO-SWS spectrum of the line of sight towards the Galactic center
(Lutz et al. \cite{Lutz96}; see Fig.~\ref{phtspec}). Observations at
our spectral resolution do however not permit a detailed comparison.
Regarding the 4.26$\mu$m CO$_2$ feature it is likely that the
feature can be attributed to solid state CO$_2$, since high spectral
resolution ISO-SWS observations of other sources indicate that the
contribution of gaseous CO$_2$ to the observed feature is very small
(see in particular van Dishoeck et al. \cite{vanDishoeck}). In the
4.6-4.8$\mu$m region, the spectra of NGC~4945 and the Galactic center
differ more strongly, and a more careful inspection is required to
assess the contributions of gaseous and solid CO and XCN. ISO-SWS
spectroscopy of the Galactic center (Lutz et al. \cite{Lutz96}) clearly
shows that what we see at low resolution as a relatively shallow and
broad feature is in fact dominated by individual lines of gaseous CO.
Contributions of a potential underlying solid CO/XCN component are
possible but difficult to separate until our high resolution follow-up
observations have been executed.

\subsection{Molecular hydrogen: physical conditions, excitation and mass}

Near infrared observations of molecular hydrogen emission in NGC~4945
have been reported by several authors over the last 15 years. The most
complete set of observations is published by Koornneef \& Israel 
(\cite{Koornneef96}), who observed 8 ro-vibrational transitions with
IRSPEC at the ESO NTT. With ISO-SWS and ISO-PHT-S we have extended 
the number of observed lines from 8 to 14 by observing the pure
rotational transitions S(0), S(1), S(2), S(3), S(5) and S(7) as well 
as the (1-0) Q(3) line. The latter was also observed with IRSPEC and 
can therefore be used to determine the proper aperture correction 
factor for the IRSPEC data set. An overview of the observed lines is 
presented in Table~\ref{molh2tab}.

\begin{table*}
\caption[]{NGC~4945 molecular hydrogen data. A($\lambda$) is the
extinction correction in magnitudes; A$_{ul}$ is the Einstein coefficient
for the transition from level u to level l. T$_{\rm u}$ is the upper level
energy of level u; g$_{\rm u}$ is the statistical weight of level u;
N$_{\rm u}$($\nu$,J) is the number of H$_2$ molecules in upper level u}
\begin{tabular}{crrccllrcl} \\ \hline
Identification & $\lambda_{\rm rest}$ & F$_{\rm obs}^{\mathrm{a}}$ & Instrument & Aperture &
 A($\lambda$)$^{\mathrm{b}}$ & A$_{\rm ul}$ & T$_{\rm u}$ & g$_{\rm u}$& 
N$_{\rm u}$($\nu$,J)/g$_{\rm u}^{\mathrm{c}}$ \\
 & $[\mu$m$]$ & $[$W/cm$^2]$ & & $[\arcsec\times\arcsec]$ & &
 $[$1/s$]$ & $[$K$]$ & & \\ \hline
0-0 S(0) & 28.2207 & 4.82$\times$10$^{-20}$ & SWS & 20$\times$27 & 0.44 & 
2.94$\times$10$^{-11}$ & 510 & 5 & 1.27$\times$10$^{62}$ \\
0-0 S(1) & 17.0346 & 1.51$\times$10$^{-19}$ & SWS & 14$\times$27 & 0.84 &
4.76$\times$10$^{-10}$ & 1015 & 21 & 5.13$\times$10$^{60}$ \\
0-0 S(2) & 12.2785 & 7.45$\times$10$^{-20}$ & SWS & 14$\times$27 & 1.04 &
2.76$\times$10$^{-9}$ & 1682 & 9 & 8.79$\times$10$^{59}$ \\ 
0-0 S(3) & 9.6649 & 5.43$\times$10$^{-20}$ & PHT-S & 24$\times$24 & 2.34 &
9.84$\times$10$^{-9}$ & 2504 & 33 & 1.28$\times$10$^{59}$ \\
0-0 S(5) & 6.9091 & 1.54$\times$10$^{-19}$ & SWS & 14$\times$20 & 0.27 &
5.88$\times$10$^{-8}$ & 4587 & 45 & 4.74$\times$10$^{57}$ \\
1-0 Q(1) & 2.4066 & 1.48$\times$10$^{-20}$ & IRSPEC & 6$\times$6 & 1.72 &
4.29$\times$10$^{-7}$ & 6150 & 9 & 9.69$\times$10$^{56}$ \\
1-0 Q(2) & 2.4134 & 5.1$\times$10$^{-21}$ & IRSPEC & 6$\times$6 & 1.71 &
3.03$\times$10$^{-7}$ & 6471 & 5 & 8.47$\times$10$^{56}$ \\
1-0 S(0) & 2.2235 & 4.4$\times$10$^{-21}$ & IRSPEC & 6$\times$6 & 1.98 &
2.53$\times$10$^{-7}$ & 6471 & 5 & 1.03$\times$10$^{57}$ \\
1-0 Q(3) & 2.4237 & 3.20$\times$10$^{-20}$ & SWS & 14$\times$20 & 1.70 &
2.78$\times$10$^{-7}$ & 6951 & 21 & 5.82$\times$10$^{56}$ \\
1-0 Q(3) & 2.4237 & 1.36$\times$10$^{-20}$ & IRSPEC & 6$\times$6 & 1.70 &
2.78$\times$10$^{-7}$ & 6951 & 21 & 5.82$\times$10$^{56}$ \\
1-0 S(1) & 2.1218 & 1.29$\times$10$^{-20}$ & IRSPEC & 6$\times$6 & 2.15 &
3.47$\times$10$^{-7}$ & 6951 & 21 & 5.84$\times$10$^{56}$ \\
0-0 S(7) & 5.5103 & 1.11$\times$10$^{-19}$ & SWS & 14$\times$20 & 0.40 &
2.00$\times$10$^{-7}$ & 7197 & 57 & 7.13$\times$10$^{56}$ \\
1-0 Q(4) & 2.4375 & 2.2$\times$10$^{-21}$ & IRSPEC & 6$\times$6 & 1.68 &
2.65$\times$10$^{-7}$ & 7584 & 9 & 2.28$\times$10$^{56}$ \\
2-1 S(1) & 2.2477 & 2.3$\times$10$^{-21}$ & IRSPEC & 6$\times$6 & 1.94 &

4.98$\times$10$^{-7}$ & 12550 & 21 & 6.36$\times$10$^{55}$\\
2-1 S(2) & 2.1542 & 0.6$\times$10$^{-21}$ & IRSPEC & 6$\times$6 & 2.09 &
5.60$\times$10$^{-7}$ & 13150 & 9 & 3.79$\times$10$^{55}$ \\ \hline
\end{tabular}
\label{molh2tab}
\begin{list}{}{}
\item[$^{\mathrm{a}}$] Before aperture correction 
\item[$^{\mathrm{b}}$] Extinction law `A' \& A$_V$=20 (see text)
\item[$^{\mathrm{c}}$] After aperture correction (IRSPEC data only)
and extinction correction. Adopted distance D=3.9Mpc
\end{list}
\end{table*}

Information on the spatial extent of the H$_2$ emitting region is only
available for the 2.12$\mu$m (1-0) S(1) line (Koornneef \& Israel
\cite{Koornneef96}; Moorwood et al. \cite{Moorwood96a}; Quillen et al. 
\cite{Quillen}; Marconi et al. \cite{Marconi}). Based on Fig.~3a of 
Moorwood et al. (\cite{Moorwood96a}) we estimate that more than 90\% 
of the (1-0) S(1) emission fits within the smallest ISO-SWS aperture 
(14$\arcsec\times$20$\arcsec$).
It is not unreasonable to expect the H$_2$ emitting area to increase 
with decreasing H$_2$ temperature. The aperture sizes used to observe 
the respective H$_2$ transitions increase with increasing sensitivity 
to lower temperature H$_2$. Based on this, we will assume in what
follows that ISO-SWS and ISO-PHT-S have observed all available warm 
H$_2$. Further to this, all three instruments were centered on the
same nuclear position and viewed the nuclear region under more or 
less similar position angles (see Sect. 2).
We will use the ratio of the 1-0 Q(3) line fluxes measured by ISO-SWS 
and IRSPEC to scale the other IRSPEC lines to the ISO-SWS aperture
size. This ratio is 2.33.

\begin{figure*}
\resizebox{\hsize}{!}{\includegraphics{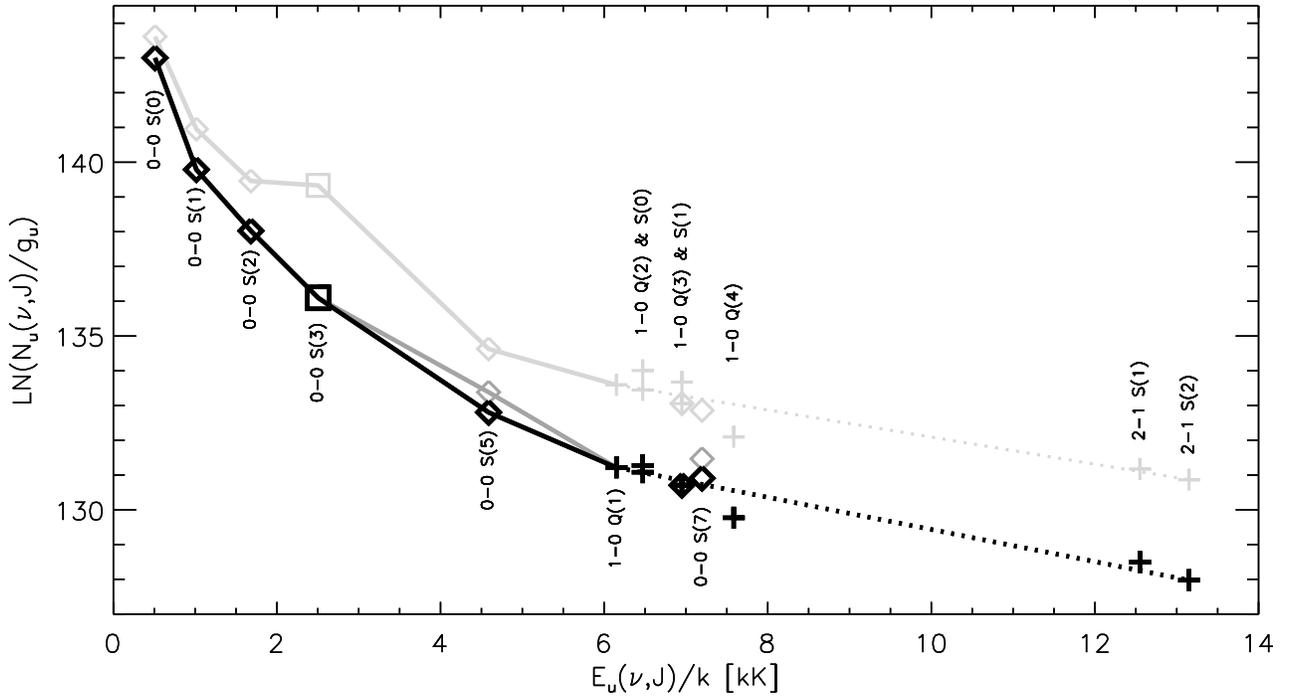}}
\caption{Excitation diagram for molecular hydrogen in
NGC~4945. Different symbols are used to distinguish different
instrumental origin: {\it diamond}: ISO-SWS; {\it square}: ISO-PHT-S;
and {\it cross}: IRSPEC. Results are shown for three different
dereddening schemes, marked by different shades of grey.
{\it Light-grey} denotes the combination extinction law `B' \&
A$_V$=50 , {\it middle-grey} denotes extinction law `B' \& A$_V$=20 and 
{\it black} extinction law `A' \& A$_V$=20}
\label{excigram}
\end{figure*}

From the 14 transitions observed it is possible to compute the upper 
level populations for 12 levels. We assumed the H$_2$ levels to be 
optically thin. The excitation diagram in Fig.~\ref{excigram} shows 
a plot of the natural logarithm of the total number of H$_2$ molecules
(N$_{\rm u}$), divided by the statistical weight (g$_{\rm u}$), in the upper level
of each transition detected, versus the energy of that level
(E$_{\rm u}$/k). The plot shows the situation after extinction correction 
(see below).

The excitation temperature (T$_{\rm ex}$) of the gas is the reciprocal of
the slope of the excitation diagram. If the warm H$_2$ is in LTE, the
excitation temperature directly corresponds to the kinetic
temperature. As is clearly visible from Fig.~\ref{excigram},
(extinction corrected; see below) the excitation temperature increases 
monotonically with upper level energy, from 160K for the combination of
(0-0) S(0) \& S(1) to 2200K for the ro-vibrational lines.

In a highly obscured galaxy like NGC~4945, extinction corrections to 
the H$_2$ emission will be important. The extinction can be estimated 
from the H$_2$ data themselves taking into account that any known 
excitation mechanism should produce a ``smooth'' excitation diagram
for the pure rotational lines, and that transitions originating in a 
common upper level should give consistent results. More specifically, 
we use three criteria:
\begin{itemize}
\item The excitation temperature should increase monotonically from 
the lowest to the highest energy levels. This sets limits on the
extinction correction for the (0-0) S(3) line in the center of the 
9.7$\mu$m silicate absorption feature.
\item The ratio of the (1-0) Q(3) \& (1-0) S(1) lines at 2.42 \& 
2.12$\mu$m should be its intrinsic ratio determined by molecular 
constants only. The same applies to the (1-0) Q(2) \& (1-0) S(0) lines
at 2.41 \& 2.22$\mu$m, that originate from identical upper levels too.
\item In LTE, the upper level populations normalized by the statistical
weights should be similar for the 0-0 S(7) \& 1-0 Q(3) lines at
5.51 \& 2.42$\mu$m, which differ by only 4\% in upper level energy.
\end{itemize}

We have varied the extinction and tried several extinction laws. 
We present the most applicable extinction laws here:
\begin{itemize}
\item Law A: A($\lambda$)$\propto\lambda^{-1.75}$ for
$\lambda<$8$\mu$m. For $\lambda>$8$\mu$m we took the Galactic center 
law of Draine (\cite{Draine}), with A(9.7$\mu$m)/E(J--K)=0.71 (Roche 
\& Aitken \cite{Roche85}) and E(J--K)=5.
\item Law B: The same as law `A', except for the range
$\lambda$=$[$2.6,8.8$]\mu$m, where we took the extinction law as 
found towards the Galactic center (Lutz \cite{Lutz99}). In the 
3-8$\mu$m range this reddening law constitutes a significantly 
higher extinction than usually assumed.
\end{itemize}

From Fig.~\ref{excigram} and the criteria defined above, moderate 
extinctions of A$_V$=17--23 are clearly preferred. Extinction law A 
provides a somewhat better fit than extinction law B. None of the 3 
solutions gives a good fit to the (1-0) Q(4) data point. In the 
following analysis, we use the preferrred extinction correction of 
A$_V$=20$^{+3}_{-3}$ and extinction law A. We note that the extinction
to the H$_2$ emitting region is slightly less than that to the starburst 
$[\ion{H}{ii}]$ regions (Sect.~3.2). This plausibly matches the 
morphological results of Moorwood et al. (\cite{Moorwood96a}), who 
find the starburst in an obscured disk, but the H$_2$ emission 
extending into a less obscured wind blown cavity.

A rough estimate of the amount of warm molecular hydrogen in the
nucleus of NGC~4945 can be derived from the level populations of
the pure rotational S(0) and S(1) transitions. The excitation 
temperature for the upper levels of these transitions (J=2 and 
J=3) is 160K. Assuming the same excitation conditions for the J=0 
and J=1 levels, we compute a warm molecular hydrogen mass of 
2.4$\times$10$^7$M$_{\sun}$. This is 9\% of the total H$_2$
gas mass estimated from CO and 0.7\% of the dynamical mass interior 
to the molecular ring (Bergman et al. \cite{Bergman}; see below).

\begin{figure*}
\resizebox{\hsize}{!}{\includegraphics{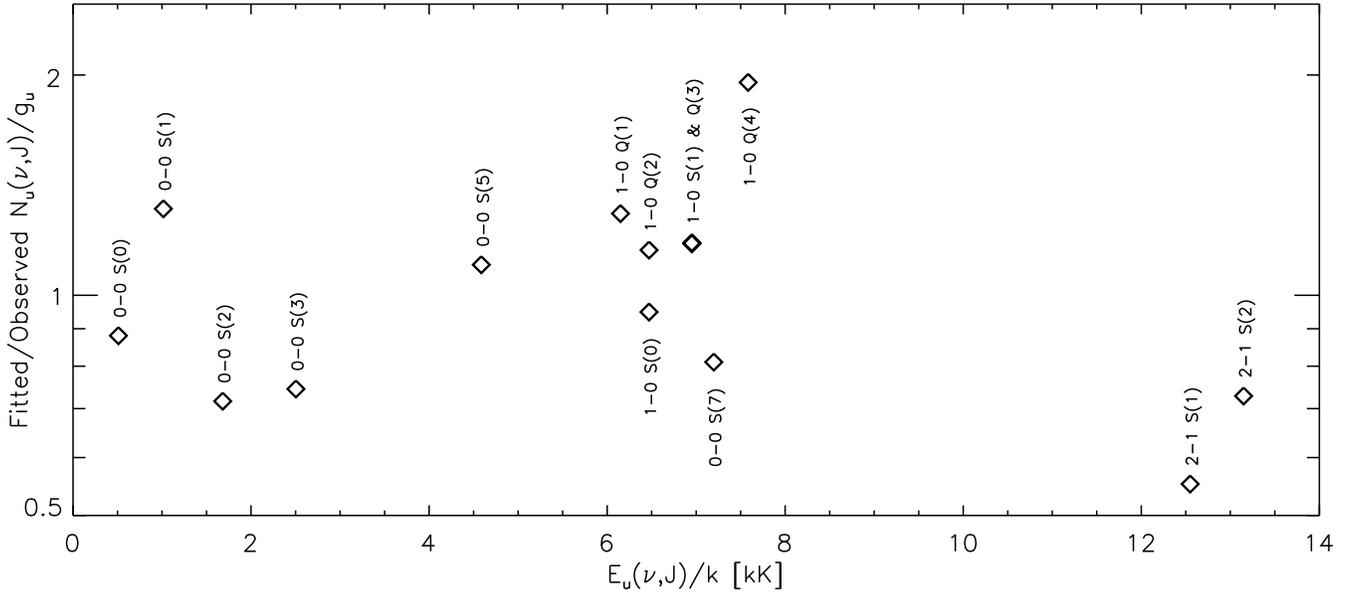}}
\caption{Ratio of the fitted and observed H$_2$ excitation diagram for
the best fitting power law dM/dT=4.43$\times$10$^{15}$ T$^{-4.793}$ M$_{\sun}$/K}
\label{excigramfit}
\end{figure*}

As already noted, the excitation temperature changes significantly
with level energy. This is the consequence of the natural fact that
the emitting gas will consist of a mixture of temperatures. The rich
NGC~4945 dataset allows us to address this in a more quantitative way.
Experiments with fits assuming a number of discrete temperature
components lead us to suggest that a power law might give a good 
representation of the mass distribution as a function of temperature.
We obtain a good fit for the following power law:
dM/dT=4.43$\times$10$^{15}$ T$^{-4.793}$ M$_{\sun}$/K. The quality 
of the fit is shown in Fig.~\ref{excigramfit}.

\begin{table}
\caption[]{Warm molecular hydrogen mass estimates for the nucleus 
of NGC~4945 using the best fit power law 
dM/dT=4.43$\times$10$^{15}$ T$^{-4.793}$ M$_{\sun}$/K. The total H$_2$ 
gas mass estimated from CO amounts to 2.7$\times$10$^8$ M$_{\sun}$}
\begin{tabular}{ccc} \\ \hline
Temp. range & M(warm H$_2$)      & \% of total M(H$_2$) \\
$[$K$]$     & $[$M$_{\sun}$$]$   &        \\ \hline
200--10000  & 2.19$\times$10$^6$ &  0.8\% \\
150--10000  & 6.51$\times$10$^6$ &  2.4\% \\
120--10000  & 1.52$\times$10$^7$ &  5.6\% \\
100--10000  & 3.03$\times$10$^7$ &   11\% \\
 90--10000  & 4.52$\times$10$^7$ &   17\% \\
 80--10000  & 7.06$\times$10$^7$ &   26\% \\
 70--10000  & 1.17$\times$10$^8$ &   43\% \\
 60--10000  & 2.10$\times$10$^8$ &   78\% \\
 50--10000  & 4.20$\times$10$^8$ &  156\% \\ \hline
\end{tabular}
\label{h2masses}
\end{table}

Table~\ref{h2masses} gives warm molecular masses for several low
temperature cut-offs. Since H$_2$ gas at temperatures below 70K does 
not contribute to the (0-0) S(0) flux, nor to any of the other pure 
rotational lines, we cannot verify whether our power law mass
distribution continues down to the lowest temperatures. Nevertheless, 
we have included mass estimates down to a low temperature cut-off of 
50K. This number is reasonable, since we don't expect the giant 
molecular clouds (GMCs) to be as cold as in the Galactic disk 
(10--20K). Rather we expect conditions as found near the Galactic 
center, where the GMCs are believed to have temperatures exceeding
50K (Armstrong \& Barrett \cite{Armstrong}).

It is interesting to compare our warm molecular hydrogen gas mass
estimate with values found in the literature (see Moorwood 
\& Oliva \cite{Moorwood94} for a review). Bergman et al. (\cite{Bergman}),
using the inner molecular rotation curve of Whiteoak et al. 
(\cite{Whiteoak}), compute a dynamical mass interior to the molecular ring
(R$\leq$280pc=15.6$\arcsec$) of 3.3$\times$10$^9$M$_{\sun}$. The same 
authors use CO to derive a total H$_2$ gas mass of 2.7$\times$10$^8$M$_{\sun}$
for the ring, assuming the rather high kinetic gas temperature of 100K. 
Note that a low temperature cut-off of the order 50--60K in our H$_2$
temperature distribution would bring our estimate of the total H$_2$ 
gas mass in agreement with that derived from the low level CO observations.
The total H$_2$ gas mass of 2.7$\times$10$^8$M$_{\sun}$ agrees well 
with a starburst-like position of NGC~4945 in the L$_{\rm IR}$--M(H$_2$) 
diagram (Young \& Scoville \cite{Young}).

In Table~\ref{h2extgal} we list for a number of external galaxies 
and Galactic template sources temperatures and masses of the warm 
molecular hydrogen gas. With a value of 9\%, NGC~4945 has a warm 
H$_2$ gas fraction similar to that found for most of the other 
galaxies listed. Note however that the results for NGC~3256, 
NGC~4038/39 and Arp~220 are less well constrained than for NGC~6946 
and NGC~4945: only for the latter two can the temperature of the 
warm H$_2$ gas be determined from the (0-0) S(0) and S(1) transitions 
directly. For the same reason a comparison of the H$_2$ gas temperatures 
is of limited value unless they are derived from identical line 
combinations. Fairly low temperatures can be derived from the 
(0-0) S(0) and S(1) lines, 160K and 179K for NGC~4945 and NGC~6946, 
respectively. Limits for other galaxies listed in Table~\ref{h2extgal}
are consistent with a similarly low temperature. This temperature is 
well below that observed for an Orion type shock ($>$500K). It is 
closer to what is observed for the same line combination in PDRs 
(e.g. Orion Bar: 155K, D. Rosenthal priv. comm.; S140: 159K, Draine 
\& Bertoldi, \cite{DraineBertoldi}).
While a variety of regions may contribute to the galaxy-integrated
temperature distribution, this comparison clearly shows Orion-like
shocks to be not representative for the entire emission, and fairly
normal PDRs (or less energetic shocks) to be perhaps more typical. If
excited by shocks (as suggested by the morphology, Moorwood et al.
\cite{Moorwood96a}), then the near-infrared H$_2$ emission in NGC~4945
must trace a small subcomponent of faster shocks.

\begin{table*}
\caption[]{Warm molecular hydrogen in external galaxies and Galactic
template sources. In column 4 detected lines are printed bold, upper 
limits normal. T$_{01}$ refers to the excitation temperature computed
from the (0-0) S(0) \& S(1) fluxes. Likewise, T$_{12}$ refers to the 
excitation temperature computed from the (0-0) S(1) \& S(2) fluxes.
T$_{\rm rot}$ is the best fit excitation temperature to several of the lowest
rotational transitions. The warm H$_2$ gas mass (M$_{\rm warm}$) is computed 
using the gas temperature listed in any of the preceeding three columns. 
The last column gives the fraction of H$_2$ gas in the warm component}
\begin{tabular}{llllcccll} \\ \hline
Object          &Type&Ref& Rot. lines observed           & T$_{01}$ & T$_{12}$ & T$_{\rm rot}$ & M$_{\rm warm}$ & \% of total H$_2$ \\
                &      & &                                          &$[$K$]$&$[$K$]$&$[$K$]$&$[$M$_{\odot}]$& \\ \hline
NGC3256         &galaxy&a& S0, {\bf S1}, {\bf S2}, {\bf S5}         &150$^{\mathrm{a}}$&   &   &10$^9$           &3\\
NGC4038/39      &galaxy&b& {\bf S1}, {\bf S2}                       &200$^{\mathrm{a}}$&   &   &5.6$\times$10$^7$&8\\
                &      & &                                          &   &405&   &8$\times$10$^6$  &1\\
NGC4945         &galaxy& & {\bf S0}---{\bf S3},{\bf S5},{\bf S7}    &160&   &   &2.4$\times$10$^7$&9\\
                &      & &                                          &   &380&   &1.2$\times$10$^6$&0.4\\
NGC6946         &galaxy&c& {\bf S0},{\bf S1},S2                     &179$^{\mathrm{b}}$&   &   &5$\times$10$^6$  &5--10\\
Arp220          &galaxy&d& S0, {\bf S1}, S2, {\bf S5}               &150$^{\mathrm{a}}$&   &   &3.5$\times$10$^9$&10\\ \hline
Orion Bar       &PDR   &e& {\bf S0}--{\bf S5},{\bf S7},{\bf S12},{\bf S13},{\bf S15},{\bf S19},{\bf S21}&155&   &   &---&---\\
S140            &PDR   &f& {\bf S0}---{\bf S5},{\bf S7},S9,{\bf S13}&159&   &   &---&---\\ \hline
Orion Peak 1    &shock &e& S0, {\bf S1}--{\bf S21}, S25             &   &578&   &---&---\\
Cepheus A West  &shock &g& {\bf S1}---{\bf S5},{\bf S7},{\bf S9}    &   &   &700&---&---\\
Cepheus A East  &shock &h& S0, {\bf S1}--{\bf S8},S9,{\bf S10},S11  &   &   &740&---&---\\ \hline
\end{tabular}
\label{h2extgal}
\begin{list}{}{}
\item[$^{\mathrm{a}}$] assumed; limits are measured for NGC~3256 ($>140$K) and
 Arp~220 ($>114$K)
\item[$^{\mathrm{b}}$] T$_{01}$ recomputed from the original (0-0) S(0) \& S(1) fluxes
\end{list}
References:\, a) Rigopoulou et al. (\cite{Rigopoulou96});\, b) Kunze et al. (\cite{Kunze});\,
c) Valentijn et al. (\cite{Valentijn96b});\, d) Sturm et al. (\cite{Sturm});\,
e) D. Rosenthal (priv. comm.);\, f) Draine\&Bertoldi (\cite{DraineBertoldi});\,
g) Wright et al. (\cite{Wright});\, h) van den Ancker et al. (\cite{vandenAncker})
\end{table*}

\section{Conclusions}

The main results of this paper can be summarized as follows:
\begin{itemize}
\item The nuclear starburst is heavily obscured by 36$^{+18}_{-11}$
mag. of visual extinction, as infered from the $[\ion{S}{iii}]$ 
18.7$\mu$m/33.5$\mu$m ratio.
\item The excitation of the nuclear starburst is very low, as deduced
from excitation indicators 
$[\ion{Ne}{iii}]$15.56$\mu$m/$[\ion{Ne}{ii}]$12.81$\mu$m and 
L$_{\rm bol}$/L$_{\rm lyc}$, consistent with an age of at least 5$\times$10$^6$yrs.
Comparison with starburst models implies that at least 50\% of the
bolometric luminosity is powered by the starburst.
\item The very low 
inferred black hole mass, the very cold mid-infrared to far-infrared colors, 
and the absence of any free line of sight to the NLR supports the conclusion 
that the starburst dominates the bolometric luminosity.
\item Our mid-infrared ISO spectroscopy does not provide any evidence
for the existence of an AGN in the nucleus of NGC~4945. 
The only high excitation line detected, the 25.9$\mu$m $[\ion{O}{iv}]$
line, is most likely produced in shocks associated with the nuclear 
starburst.
\item The AGN, detected in hard X-rays, is unusual in not revealing 
itself at optical, near-infrared and mid-infrared wavelengths. 
Hence, either the NLR is extremely obscured (A$_V>$160), 
or UV photons from the AGN are absorbed close to the nucleus
along all lines of sight, or the AGN is deficient in UV relative to
its X-ray flux.
\item Many ISM solid state and molecular features have been observed
with ISO-PHT-S in the 2.4--11.7$\mu$m range. Most prominent in
emission are the PAH features at 3.3, 6.2, 7.7 and 11.2$\mu$m.
The strongest absorption features are those of water ice, CO$_2$ and 
CO, seen against the nuclear spectrum. These features show
striking similarities to the absorption features seen towards the 
Galactic center.
\item We have studied the physical conditions, excitation and mass of
warm H$_2$, combining IRSPEC and ISO observations of 14 transitions. 
We derive a visual extinction of 20$^{+3}_{-3}$ mag. to the H$_2$ 
emitting region. From the (0-0) S(0)\& S(1) lines, we compute a warm 
(160K) H$_2$ gas mass of 2.4$\times$10$^7$M$_{\odot}$, 9\% of the 
total gas mass inferred from CO. The excitation diagram is best fitted
by a power law of the form dM/dT=4.43$\times$10$^{15}$ T$^{-4.793}$ 
M$_{\sun}$/K. The low excitation temperature of 160K shows Orion-like 
shocks not to be representative for the entire emission, and fairly 
normal PDRs to be perhaps more typical.
\end{itemize}

\begin{acknowledgements}
The authors wish to thank Dietmar Kunze and Fred Lahuis for help in
the SWS datareduction and Matt Lehnert, Steve Lord and Eckhard Sturm 
for stimulating discussions.
\end{acknowledgements}


\begin{thebibliography}{}

\bibitem[1987]{Ables}Ables J.G., Forster J.R., Manchester R.N., et
        al., 1987, MNRAS 226, 157
\bibitem[1996]{Acosta96}Acosta-Pulido J.A., Klaas U., Laureijs R.J., 
        et al., 1996, A\&A 315, L121
\bibitem[1999]{Acosta99}Acosta-Pulido J.A., 1999, ISO Explanatory Library, 
        http://www.iso.vilspa.esa.es/users/expl\_lib/PHT/chop\_re-
        port02.ps.gz
\bibitem[1985]{Armstrong}Armstrong J.T., Barrett A.H., 1985, ApJS 57, 535
\bibitem[1992]{Bergman}Bergman P., Aalto S., Black J.H., Rydbeck G., 
        1992, A\&A 265, 403
\bibitem[1988]{Brock}Brock D., Joy M., Lester D.F., et al., 1988,
        ApJ 329, 208
\bibitem[2000]{Chiar}Chiar J.E., Tielens A.G.G.M., Whittet D.C.B.,
        et al., 2000, ApJ in press (astroph/0002421)
\bibitem[2000]{Clavel}Clavel J., Schulz B., Altieri B., et al.,
        2000, A\&A in press (astroph/0003298)
\bibitem[1996]{deGraauw}De Graauw Th., Haser L.N., Beintema D.A., 
        et al., 1996, A\&A 315, L49
\bibitem[1989]{Draine}Draine B.T., 1989, In: Infrared Spectroscopy in
        Astronomy, B.H. Kaldeich (ed.), ESA-SP 290, 93
\bibitem[1999]{DraineBertoldi}Draine B.T., Bertoldi F., 1999, 
        In: The Universe as seen by ISO, P. Cox, M.F. Kessler (eds.),
        ESA-SP 427, 553
\bibitem[1994]{Elvis}Elvis M., Wilkes B., McDowell J.C., et al., 1994,
        ApJS 95, 1
\bibitem[1998]{Forbes}Forbes D.A., Norris R.P., 1998, MNRAS 300, 757
\bibitem[1997]{Gabriel}Gabriel C., Acosta-Pulido J., Heinrichsen I.,
        Morris H., Tai W.-M., 1997, In: Proc. of the ADASS VI
        conference, G. Hunt \& H.E. Payne (eds.), ASP Conf. Ser. 125, 108
\bibitem[1998]{Genzel}Genzel R., Lutz D., Sturm E., et al., 1998, 
        ApJ 498, 579
\bibitem[1995]{Gerakines95}Gerakines P.A., Schutte W.A., Greenberg J.M., 
        van Dishoeck E.F., 1995, A\&A 296, 810
\bibitem[1999]{Gerakines99}Gerakines P.A., Whittet D.C.B.,
        Ehrenfreund P., et al., 1999, ApJ 522, 357
\bibitem[1997]{Greenhill}Greenhill L.J., Moran J.M., Herrnstein J.R.,
        1997, ApJ 481, L23
\bibitem[2000]{Guainazzi}Guainazzi M., Matt G., Brandt W.N., et
        al., 2000, A\&A in press (astroph/0001528) 
\bibitem[1981]{Hagen}Hagen W., Tielens A.G.G.M., 1981,
        J. Chem. Phys. 75, 4198
\bibitem[1990]{Heckman}Heckman T.M., Armus L., Miley G.K., 1990,
        ApJS 74, 833
\bibitem[1984]{Hesser}Hesser J.E., Harris H.C., van den Bergh S.,
        et al., 1984, ApJ 276, 491
\bibitem[1993]{Iwasawa}Iwasawa K., Koyama K., Awaki H., et al.,
        1993, ApJ 409, 155
\bibitem[1996]{Kessler}Kessler M.F., Steinz J.A., Anderegg M.E., 
        et al., 1996, A\&A 315, L27
\bibitem[1993]{Koornneef93}Koornneef J., 1993, ApJ 403, 581
\bibitem[1996]{Koornneef96}Koornneef J., Israel P.F., 1996, New
        Astronomy 1, 271
\bibitem[1996]{Kunze}Kunze D., Rigopoulou D., Lutz D., et al.,
        1996, A\&A 315, L101
\bibitem[1984]{Lacy}Lacy J.H., Baas F., Allamandola L.J., 1984, ApJ 276, 533
\bibitem[1998]{Lahuis}Lahuis F., Wieprecht E., Bauer O.H., et al., 1998, In:
        Astronomical Data Analysis Software and Systems VII, R. Albrecht, 
        R.N. Hook, H.A. Bushouse (eds.), ASP Conf. Ser. 145, 224
\bibitem[1995]{Leitherer}Leitherer C., Heckman T.M., 1995, ApJS 96, 9
\bibitem[1996]{Lemke}Lemke D., Klaas U., Abolins J., et al., 1996, A\&A 315, 
        L64
\bibitem[1999]{Lutz99}Lutz D., 1999, In: The Universe as seen by ISO,
        P. Cox, M.F. Kessler (eds.), ESA-SP 427, 623
\bibitem[1996]{Lutz96}Lutz D., Feuchtgruber H., Genzel R., et al., 1996, A\&A 
        315, L269
\bibitem[1998]{Lutz98}Lutz D., Kunze D., Spoon H.W.W., et al., 1998, A\&A 
        333, L75
\bibitem[2000]{Marconi}Marconi A., Maiolino R., Oliva E., et al., 2000,
        A\&A in press (astroph/0002244)
\bibitem[1999]{Matt}Matt G., Guainazzi M., Maiolino R., et al.,
        1999, A\&A 341, L39
\bibitem[1999]{Mattila}Mattila K., Lehtinen K., Lemke D., 1999, A\&A 342, 643
\bibitem[1996]{Mauersberger}Mauersberger R., Henkel C., Whiteoak J.B.,
        et al., 1996, A\&A 309, 705
\bibitem[1984]{Moorwood84}Moorwood A.F.M., Glass I.S., 1984, A\&A 135, 281
\bibitem[1988]{Moorwood88}Moorwood A.F.M., Oliva E., 1988, A\&A 203, 278
\bibitem[1994]{Moorwood94}Moorwood A.F.M., Oliva E., 1994, ApJ 429, 602
\bibitem[1996a]{Moorwood96a}Moorwood A.F.M.,van der Werf P.P.,
        Kotilainen J.K., et al., 1996a, A\&A 308, L1
\bibitem[1996b]{Moorwood96b}Moorwood A.F.M., Lutz D., Oliva E., et al.,
        1996b, A\&A 315, L109
\bibitem[1999]{Oliva}Oliva E., Moorwood A.F.M., Drapatz S., et al.,
        1999, A\&A 343, 943
\bibitem[1995]{Ott}Ott M., 1995, Ph.D. Thesis, Bonn University
\bibitem[1999]{Quillen}Quillen A.C., Alonso-Herrero A., Rieke M.J.,
        et al., 1999, ApJ 527, 696
\bibitem[1988]{Rice}Rice W., Lonsdale C.J., Soifer B.T., et al.,
        1988, ApJS 68, 91
\bibitem[1996]{Rigopoulou96}Rigopoulou D., Lutz D., Genzel R., 
        et al., 1996, A\&A 315, L125
\bibitem[1999]{Rigopoulou99}Rigopoulou D., Spoon H.W.W., Genzel R., et al., 
        1999, AJ 118, 2625
\bibitem[1999]{Risaliti}Risaliti G., Maiolino R., Salvati M., 1999,
        ApJ 522, 157
\bibitem[1985]{Roche85}Roche P.F., Aitken D.K., 1985, MNRAS 215, 425
\bibitem[1995]{Sadler}Sadler E.M., Slee O.B., Reynolds J.E., Roy A.L.,
        1995, MNRAS 276, 1373 
\bibitem[1996]{Schaeidt}Schaeidt S., Morris P.W., Salama A., et al., 1996,
        A\&A 315, L55
\bibitem[1997]{Siebenmorgen}Siebenmorgen R., Moorwood A., Freudling W., 
        et al., 1997, A\&A 325, 450
\bibitem[1996]{Sturm}Sturm E., Lutz D., Genzel R., et al., 1996,
        A\&A 315, L133
\bibitem[1999]{Tayal}Tayal S.S., Gupta G.P., 1999, ApJ 526, 544
\bibitem[2000]{Thornley}Thornley M.D., F\"{o}rster Schreiber N.M.,
        Lutz D., et al., 2000, ApJ in press (astroph/0003334)
\bibitem[1996a]{Valentijn96a}Valentijn E., Feuchtgruber H., Kester
        D.J.M., et al., 1996, A\&A 315, L60
\bibitem[1996b]{Valentijn96b}Valentijn E., Van der Werf P.P., 
        De Graauw Th., et al., 1996, A\&A 315, L145
\bibitem[2000]{vandenAncker}van den Ancker M.E., Tielens A.G.G.M.,
        Wesselius P.R., 2000, A\&A in prep.
\bibitem[1996]{vanDishoeck}van Dishoeck E.F., Helmich F.P., de
        Graauw Th., et al., 1996, A\&A 315, L349
\bibitem[1990]{Whiteoak}Whiteoak J.B., Dahlem M., Wielebinski R.,
        Harnett J.I., 1990, A\&A 231, 25
\bibitem[1998]{Wieprecht}Wieprecht E., Lahuis F., Bauer O.H., et al.,
        1998, In: Astronomical Data Analysis Software and Systems VII, 
        R. Albrecht, R.N. Hook, H.A. Bushouse (eds.), ASP Conf. Ser. 145, 279
\bibitem[1996]{Wright}Wright C.M., Drapatz S., Timmermann R., 
        et al., 1996, A\&A 315, L301
\bibitem[1991]{Young}Young J.S., Scoville N., 1991, ARA\&A 29, 581

\end{thebibliography}
\end{document}